# ELECTROWEAK SYMMETRY BREAKING AND PHYSICS BEYOND THE STANDARD MODEL

SALLY DAWSON

*Physics Department, Brookhaven National Laboratory, Upton, N.Y. 11973*

HOWARD E. HABER

*Santa Cruz Institute for Particle Physics*
*University of California, Santa Cruz, CA 95064*

In order to extend the Standard Model to TeV scale energies one must address two basic questions: (1) What is the complete description of the effective theory of fundamental particles at and below the electroweak scale? and (2) What is the dynamics responsible for electroweak symmetry breaking? The answers to these questions are crucial for addressing the third outstanding question of particle physics: What are the origins of the Standard Model parameters? We briefly summarize current theoretical approaches to answering some of these questions.

## 1 Overview

The development of the Standard Model of particle physics is a remarkable success story. Its many facets have been tested at present day accelerators and no unambiguous deviations have been found. In some cases, the model has been verified at an accuracy of better than one part in a thousand. This state of affairs presents our field with a challenge. Where do we go from here?

Despite the success of the Standard Model as a description of the properties of fundamental particles and interactions, the Standard Model cannot be a fundamental theory; at best it is a low-energy approximation to a more fundamental theory of particle interactions. In order to extend the Standard Model to higher energies, one must address three basic questions: (i) what is the complete description of the effective theory of fundamental particles at and below the electroweak scale? (ii) what is the mechanism for electroweak symmetry breaking? and (iii) what are the origins of the Standard Model parameters?

At present, the dynamics underlying the electroweak symmetry breaking mechanism are unknown. However, theoretical studies lead to one very important conclusion: *the nature of the dynamics of electroweak symmetry breaking must be revealed at or below the 1 TeV energy scale.* The most conservative approach would say that the study of the TeV scale must reveal the Higgs boson, but with no guarantees of anything beyond. However, strong theoretical arguments suggest that the dynamics underlying electroweak symmetry breaking must be richer. In particular, a comprehensive experimental study of the TeV scale is likely to reveal a new sector of physics beyond the Standard Model (for example, a super-particle spectrum of low-energy supersymmetry or a new strongly-interacting sector of fermions, scalars, and/or vector resonances). *The critical path to maintain the long-term vitality of particle physics must therefore involve the thorough exploration of the TeV-scale in order to elucidate the dynamics of electroweak symmetry breaking and complete the "low-energy" description of particle physics.*

The origin of the Standard Model parameters presents a challenge which may or may not be addressed by the next generation of colliders. This is because the energy scale associated with flavor physics (which controls most of the Standard Model parameters) is not constrained by present theories or experimental data, and could lie far beyond the energy scale accessible to future colliders. Nevertheless, the elucidation of TeV-scale physics may have significant ramifications for our understanding of flavor and physics at even higher energies.

This book focuses on electroweak symmetry breaking and examines the potential for probing models of physics beyond the Standard Model at future accelerators. (A summary of this work was first presented in Ref. [1].) Exploration of TeV-scale physics requires a new generation of colliders and detectors beyond those currently in operation or construction. The capabilities of the approved future facilities and a variety of hypothetical future machines were considered; these machines are summarized in Table 1.

Table 1:

| Name | Type | $\sqrt{s}$ | Yearly $\int \mathcal{L} dt$ |
|---|---|---|---|
| *Approved Projects* | | | |
| LEP-2 | $e^+e^-$ | 160–192 GeV | 100–200 pb$^{-1}$ |
| Tevatron (Main Injector) | $p\bar{p}$ | 2 TeV | 1 fb$^{-1}$ |
| LHC | $pp$ | 14 TeV | 10–100 fb$^{-1}$ |
| *Tevatron Upgrades (hypothetical)* | | | |
| TeV* | $p\bar{p}$ | 2 TeV | 10 fb$^{-1}$ |
| Di-Tevatron | $p\bar{p}$ | 4 TeV | 20 fb$^{-1}$ |
| $e^+e^-$ *Linear Collider* | | | |
| JLC, NLC, TESLA | $e^+e^-$ † | 0.5–1.5 TeV | 50–200 fb$^{-1}$ |
| | († with $e\gamma, \gamma\gamma, e^-e^-$ options) | | |
| *Other Futuristic Colliders* | | | |
| $\mu^+\mu^-$ Collider | $\mu^+\mu^-$ | 0.5–4 TeV | 20–1000 fb$^{-1}$ |
| Hadron supercollider | $pp$ | 60 TeV | 100 fb$^{-1}$ |

The origin of the Standard Model parameters is not directly addressed in this book. Nor does this book examine the possibility that new and interesting physics beyond the Standard Model might populate the energy interval *above* the TeV-scale. The Standard Model does *not* provide any clues as to the energy scale responsible for addressing the origin of the Standard Model parameters (or a supersymmetric extension thereof). The relevant energy scale could lie anywhere from 1 TeV to the Planck scale, $M_P$. For example, in conventional theories of low-energy supersymmetry, the origin of the Standard Model and supersymmetric parameters lies at or near the Planck scale. In contrast, theories of extended technicolor typically attempt to solve the flavor problem at energy scales below 100 TeV. This book focuses directly on the TeV-scale physics accessible to the next generation of colliders. Namely, the requirements for completing the TeV-scale particle spectrum and unraveling the dynamics of electroweak symmetry breaking are examined. We expect that future colliders will discover new physics beyond the Standard Model associated with this dynamics.

In Section 2, we begin with a brief review of the Standard Model [2]. While the Standard Model is extremely successful in describing the experimental data, there are a number of theoretical objections associated with the mechanism of electroweak symmetry breaking. The Standard Model assumes that electroweak symmetry breaking is the result of the dynamics of an elementary complex doublet of scalar fields. The neutral component of the scalar doublet acquires a vacuum expectation value and triggers electroweak symmetry breaking; the resulting particle spectrum contains three massive gauge bosons and the (massless) photon, massive quarks and charged leptons, massless neutrinos, and a massive Higgs scalar. However, if one attempts to embed the Standard Model in a fundamental theory with a high energy scale (such as the Planck scale, $M_P$), then the masses of elementary scalars naturally assume values of order the high energy scale. This is due to the quadratic sensitivity of squared scalar masses to the highest energy scale of the underlying fundamental theory. In contrast, theoretical considerations (*e.g.* unitarity) require the mass of the Standard Model Higgs boson to be of order the electroweak scale. To accomplish this in the Standard Model, one must "unnaturally" adjust the parameters of the high-energy theory (a fine tuning of 34 orders of magnitude in the scalar squared mass parameter is required) to produce the required light elementary scalar field. These and other theoretical shortcomings of the Standard Model are outlined in Section 3.

Overcoming these theoretical deficiencies requires new physics beyond the Standard Model which controls the dynamics of the electroweak symmetry breaking. The generic properties of the new physics that can be obtained independently of specific models are described in Section 4. In particular, experimental evidence for the dynamics underlying electroweak symmetry breaking must be revealed at the TeV energy scale (or below). However, to make significant progress, one must introduce a specific model framework. Two main approaches have been proposed:

(i) Invoke supersymmetry to cancel quadratic sensitivity of the scalar squared masses to $M_P$. Electroweak symmetry breaking is triggered by the dynamics of a weakly-coupled Higgs sector.

(ii) Elementary Higgs scalars do not exist. A Higgs-like scalar state (if it exists) would reveal its composite nature at the TeV-scale, where new physics beyond the Standard Model enters. Electroweak symmetry breaking is triggered by non-perturbative strong forces.

In order to implement case (i), one must first note that supersymmetry is not an exact symmetry of nature (otherwise, all known particles would have equal-mass supersymmetric partners). If supersymmetry is to explain why the Higgs boson mass is of order the electroweak scale, then supersymmetry-breaking effects which split the masses of particles and their super-partners must be roughly of the same order as the electroweak scale. Thus, if supersymmetry is connected with the origin of electroweak symmetry breaking, one expects to discover a new spectrum of particles and interactions at the TeV-scale or below. In addition, such models of "low-energy" supersymmetry are compatible with the existence of weakly-coupled elementary Higgs scalars with masses of order the electroweak symmetry breaking-scale. A comprehensive treatment of Higgs boson phenomenology is given in Chapter 2.

No direct experimental evidence for the supersymmetric particle spectrum presently exists. There is, however, tantalizing indirect evidence. Starting from the known values of the SU(3), SU(2), and U(1) gauge couplings at $m_Z$ and extrapolating to high energies, one finds that the three gauge couplings meet at a single point if one includes the effects of supersymmetric particles (with masses at or below 1 TeV) in the running of the couplings. Unification of couplings then takes place at around $10^{16}$ GeV, only a few orders of magnitude below $M_P$. The three gauge couplings do not meet at a single point if only Standard Model particles contribute to the running of the couplings. This could be a hint for low-energy supersymmetry and suggests that the theory of fundamental particles remains weakly coupled and perturbative all the way up to energies near $M_P$. This approach is discussed in Section 5.

New physics is also invoked to explain the origin of electroweak symmetry breaking in case (ii). For example, in technicolor models (which make use of the mechanism analogous to the one that is responsible for chiral symmetry breaking in QCD), electroweak symmetry breaking occurs when pairs of techni-fermions condense in the vacuum. One then identifies a new scale, $\Lambda_{ESB} \simeq 4\pi v \sim \mathcal{O}(1 \text{ TeV})$, where new physics beyond the Standard Model must enter. Other approaches, such as effective Lagrangian descriptions of the strongly interacting Higgs sector, preon models, top-mode condensate models, composite Higgs models, *etc.*, also fall into this category. Unfortunately, due to the presence of non-perturbative strong forces, it is often difficult to make reliable detailed computations in such models. Moreover, phenomenological difficulties inherent in the simplest examples require additional structure (*e.g.*, an extended technicolor sector is needed in technicolor models to generate fermion masses). This approach is discussed in Section 6. Unfortunately, a completely phenomenologically viable fundamental model of strongly-coupled electroweak symmetry breaking has not yet been constructed.

Both supersymmetric and technicolor models contain a full spectrum of new particles which may be accessible at the next generation of colliders. The phenomenology of these particles play a key role in the search for new phenomena and is addressed in detail in Chapters 3–6. Nevertheless, these visions may be too limited. Additional unexpected new particles and interactions might accompany the physics underlying electroweak symmetry breaking. Clearly, in order to extrapolate our particle theories to higher energies with any confidence, we must know the complete TeV-scale spectrum. The Standard Model possesses three generations of quarks and leptons (and no right-handed neutrinos) and an SU(2)×U(1) gauge group. Is that all? Might there be additional particles that populate the TeV scale? Some possibilities considered in Chapters 7 and 8 include: (i) an extended electroweak gauge group yielding additional gauge bosons, (*e.g.*, extra U(1) factors, left-right (LR) symmetry such as $SU(2)_L \times SU(2)_R$, *etc.*); (ii) new fermions (*e.g.*, a fourth generation, mirror fermions which possess right-handed couplings, vector-like fermions, massive neutrinos, *etc.*); (iii) new bosons (extended Higgs sectors, pseudo-Goldstone bosons, other exotic scalars, leptoquarks, vector resonances, *etc.*). New TeV scale physics can also be detected through virtual interactions of new particles that contribute to anomalous couplings and other low-energy observables, as reviewed in Chapters 9 and 10.

Finally, we briefly summarize in Section 7 the key questions of TeV-scale particle physics. One of the major goals of this book is to examine how best to answer these questions at the next generation of colliders. Detector and accelerator issues relevant to the search for TeV-scale physics at future hadron and $e^+e^-$ colliders are examined in Chapters 11 and 12. The elucidation of the TeV-scale will surely bring forth a new era in the development of the fundamental theory of particles and their interactions.

## 2  The Standard Model of Electroweak Interactions

The electroweak sector of the Standard Model is an $SU(2)_L \times U(1)_Y$ gauge theory. The bosonic sector of the theory contains three massless $SU(2)_L$ gauge bosons, $W^i_\mu$, one massless $U(1)_Y$ gauge boson, $B_\mu$, and a hypercharge 1, $SU(2)_L$ doublet of complex scalar fields,
$$\Phi \equiv \frac{1}{\sqrt{2}} \begin{pmatrix} \phi_1 + i\phi_2 \\ H + i\phi_3 \end{pmatrix}. \tag{2.1}$$

The Lagrangian is given by
$$\mathcal{L} = -\tfrac{1}{4}W^i_{\mu\nu}W^{\mu\nu i} - \tfrac{1}{4}B_{\mu\nu}B^{\mu\nu} + (D_\mu \Phi)^\dagger(D^\mu \Phi) - V(\Phi), \tag{2.2}$$

where
$$\begin{aligned}
W^i_{\mu\nu} &= \partial_\mu W^i_\nu - \partial_\nu W^i_\mu - g\epsilon^{ijk}W^j_\mu W^k_\nu \\
B_{\mu\nu} &= \partial_\mu B_\nu - \partial_\nu B_\mu \\
D_\mu &= \partial_\mu + \tfrac{1}{2}ig\tau^i W^i_\mu + \tfrac{1}{2}ig'Y B_\mu \\
V(\Phi) &= \mu^2 \mid \Phi^\dagger \Phi \mid + \lambda \mid \Phi^\dagger \Phi \mid^2 .
\end{aligned} \tag{2.3}$$

The theory depends on the following free parameters: the $SU(2)_L$ and $U(1)_Y$ gauge coupling constants ($g$ and $g'$) and the scalar mass and quartic coupling parameters ($\mu^2$ and $\lambda$), which appear in the scalar potential $V$. Explicit mass terms for the gauge bosons are forbidden by the gauge invariance. Eq. (2.2) is the most general renormalizable and $SU(2)_L \times U(1)_Y$ invariant Lagrangian allowed involving only the gauge bosons and scalar fields.

The state of minimum energy for $\lambda > 0$ and $\mu^2 < 0$ is not at zero and the scalar field develops a vacuum expectation value (VEV) which can be taken to be

$$\langle \Phi \rangle = \frac{1}{\sqrt{2}} \begin{pmatrix} 0 \\ v \end{pmatrix}. \tag{2.4}$$

The symmetry breaking scheme exhibited is $SU(2)_L \times U(1)_Y \to U(1)_{EM}$. In unitary gauge we can write the scalar field as

$$\Phi = \frac{1}{\sqrt{2}} \begin{pmatrix} 0 \\ H + v \end{pmatrix}. \tag{2.5}$$

The field $H$ is the Higgs boson. It is a physical scalar which can be produced and detected experimentally. It is straightforward to see that the kinetic energy term for $\Phi$ now generates mass terms for three of the gauge bosons, $W^\pm$ and $Z$, while the fourth gauge boson, the photon, remains massless. This is a simple realization of the Higgs mechanism, in which the electroweak symmetry is spontaneously broken, while maintaining renormalizability and unitarity of the theory. Finding the physical remnant of this mechanism, the Higgs boson, is a vital test of the correctness of the model.

The Higgs mechanism is also used to generate quark and (charged) lepton masses. The $SU(2)_L \times U(1)_Y$ gauge invariant Yukawa couplings of the Higgs boson to "up-type" and "down-type" fermions are

$$\mathcal{L}_f = -\lambda_d \overline{Q}_L \Phi d_R - \lambda_u \overline{Q}_L \Phi^c u_R + \text{h.c.} \tag{2.6}$$

where $\Phi^c \equiv i\tau_2 \Phi^*$ transforms as a hypercharge $-1$ $SU(2)_L$ doublet and $\overline{Q}_L \equiv (\overline{u}_L, \overline{d}_L)$. When $\Phi$ is replaced by its VEV, Eq. (2.6) generates mass terms for the fermions, with

$$\lambda_f = \frac{m_f \sqrt{2}}{v}. \tag{2.7}$$

Note that within the Standard Model, there is no explanation for the value of the fermion masses. They are simply input parameters of the theory.

One of the most important properties of the Higgs boson is that its couplings to fermions and gauge bosons are proportional to the corresponding particle masses. Moreover, the overall strength of these couplings is determined in terms of known masses and couplings. The scalar potential of Eq. (2.3) initially has two free parameters, $\mu$ and $\lambda$. We can trade these in for

$$\begin{aligned} v^2 &= -\frac{\mu^2}{2\lambda} \\ M_H^2 &= 2v^2\lambda. \end{aligned} \tag{2.8}$$

The muon decay rate, $\Gamma(\mu \to \nu_\mu e \overline{\nu}_e)$, determines $v^2 = (\sqrt{2} G_F)^{-1} = (246 \text{ GeV})^2$. Thus, *the only remaining unknown parameter is the Higgs mass.* As a result, the Standard Model Higgs boson production and decay rates can be computed unambiguously in terms of the Higgs boson mass. Chapter 2 of this book discusses the search for the Standard Model Higgs boson at future colliders.

From Eq. (2.8), we see that as the Higgs boson becomes heavy, the quartic coupling $\lambda$ in the scalar potential becomes large; for $\lambda \gtrsim \mathcal{O}(1)$ we say that the scalar sector of the theory is strongly coupled. The physics of a strongly interacting electroweak symmetry breaking sector is discussed in Chapter 5.

Despite the wealth of precision electroweak data, at present we have only very weak experimental limits on the Higgs boson mass. From direct searches at LEP we have the limit [3]

$$M_H > 65.2 \text{ GeV}. \tag{2.9}$$

An improved bound of around 95 GeV is achievable if no Higgs boson is discovered at LEP-2 [4]. Standard Model Higgs boson searches at the LHC will be sensitive to masses up to around 700 GeV.

Precision measurements at LEP give an indirect limit on the Higgs boson mass from loop effects in electroweak radiative corrections [5],

$$M_H < 300 \text{ GeV} \quad (95\% \text{ CL}). \tag{2.10}$$

This limit, however, depends sensitively on which pieces of experimental data are included in the fit and assumes the correctness of the minimal Standard Model.

(For example, if the measurements of the $Z \to b\bar{b}$ and $Z \to c\bar{c}$ decay rates are omitted from the fit, the 95% CL upper bound from Ref. [5] becomes $M_H < 450$ GeV.) Loop effects which involve the Higgs boson in a number of low energy processes are discussed in Chapter 11, although these are relevant mostly in extensions of the Standard Model with non-minimal Higgs sectors.

## 3 The Standard Model is not Complete

Despite the success of the Standard Model as a description of the properties of fundamental particles, it is clear that the Standard Model cannot be the fundamental theory of nature. There are numerous theoretical objections to the Standard Model which we briefly summarize here. However, it should be emphasized that the validity of the Standard Model at all energy scales below the Planck scale is not presently contradicted by any *confirmed* experimental data.

For simplicity consider the Higgs sector of the Standard Model (with gauge bosons and fermions omitted); *i.e.*, a pure scalar theory in which the potential $V(\Phi)$ is given by Eq. (2.3). The quartic coupling $\lambda$ runs with the renormalization scale $Q$:

$$\frac{d\lambda}{dt} = \frac{3\lambda^2}{4\pi^2}, \tag{3.1}$$

where $t \equiv \log(Q^2/Q_0^2)$ and $Q_0$ is some reference scale. This equation is easily solved:

$$\frac{1}{\lambda(Q)} = \frac{1}{\lambda(Q_0)} - \frac{3}{4\pi^2} \log\left(\frac{Q^2}{Q_0^2}\right), \tag{3.2}$$

or equivalently,

$$\lambda(Q) = \frac{\lambda(Q_0)}{1 - [3\lambda(Q_0)/4\pi^2] \log(Q^2/Q_0^2)}. \tag{3.3}$$

We see that $\lambda(Q)$ blows up as $Q \to \infty$ (with $\lambda(Q_0)$ fixed and positive). Regardless of how small $\lambda(Q_0)$ is, $\lambda(Q)$ will eventually become infinite at some large $Q$, called the Landau pole. Alternatively, with $\lambda(Q_0) > 0$, $\lambda(Q) \to 0$ as $Q \to 0$; *i.e.*, the theory, if valid to arbitrarily high energy scales, becomes a non-interacting theory at low energy, a so-called trivial theory [6]. To avoid this theoretical embarrassment, one typically attempts to choose parameters in such a way that the Landau pole lies above the Planck scale. In that case, one can hope that a more fundamental Planck scale theory avoids this problem. If the Landau pole lies below the Planck scale, one must invoke additional interactions (which would either alter the renormalization flow of the coupling or fundamentally change the degrees of freedom of the theory) at some intermediate energy scale near or below the predicted location of the Landau pole. The inclusion of gauge boson and fermion fields does not alter these conclusions. In order to avoid a Landau pole below the Planck scale, the Higgs boson of the Standard Model must be relatively light, with mass less than about 180 GeV [7]. This is sometimes interpreted as an upper bound on the Standard Model Higgs boson mass.

If no Higgs boson is discovered with a mass lying below 180 GeV, one would conclude that there must be a new energy scale in particle physics that lies below the Planck scale. Still, if this new scale were much larger than 1 TeV, it would have no discernible effect on experimental observables at present colliders (as well as at colliders of the foreseeable future).

There have been many other attempts to limit the allowed mass region of the Higgs boson and to pinpoint the energy scale at which either the Higgs boson or new physics must be revealed. Another simple bound is based on perturbative unitarity [8]. The $J = 0$ partial wave for the elastic scattering of longitudinal $W$ bosons in the high energy limit, $s \gg m_W^2$, is

$$a_0(W_L^+ W_L^- \to W_L^+ W_L^-) = -\frac{G_F M_H^2}{8\sqrt{2}\pi}\left[2 + \frac{M_H^2}{s - M_H^2} - \frac{M_H^2}{s}\log\left(1 + \frac{s}{M_H^2}\right)\right]. \quad (3.4)$$

At very high energies, $s \gg M_H^2$, one finds

$$a_0(W_L^+ W_L^- \to W_L^+ W_L^-) \longrightarrow -\frac{G_F M_H^2}{4\pi\sqrt{2}}. \quad (3.5)$$

Applying the condition for perturbative unitarity, $|\mathrm{Re}\, a_0| < \frac{1}{2}$, gives the restriction $M_H < 870$ GeV. The most restrictive bound is derived from a coupled channel analysis and gives

$$M_H^2 < \frac{4\pi\sqrt{2}}{3G_F} \simeq (700 \text{ GeV})^2. \quad (3.6)$$

Similar bounds are found from lattice studies [9]. It is important to understand that this does not mean that the Higgs boson cannot be heavier than 700 GeV, it simply means that for heavier masses perturbation theory is not valid and the electroweak sector of the theory is strongly interacting.

We can apply an alternate limit to Eq. (3.4) and take the Higgs boson much heavier than the energy $\sqrt{s}$. For $M_H^2 \gg s$, one finds

$$a_0(W_L^+ W_L^- \to W_L^+ W_L^-) \longrightarrow \frac{G_F s}{16\pi\sqrt{2}}. \quad (3.7)$$

Again, applying the perturbative unitarity condition one finds $\sqrt{s_c} < 1.8$ TeV, where we have used the notation $\sqrt{s_c}$ to denote the critical energy scale at which perturbative unitarity is violated. In this case, the most restrictive bound emerges from considering the isospin zero channel and yields

$$s \leq s_c \equiv \frac{4\pi\sqrt{2}}{G_F} \lesssim (1.2 \text{ TeV})^2. \quad (3.8)$$

Eq. (3.8) is the basis for the often repeated assertion that *there must be new physics at the TeV scale.* Eq. (3.8) is telling us that without a Higgs boson, there must be new physics that restores perturbative unitarity somewhere below an energy scale of around 1.2 TeV. This is precisely the energy scale that will be probed at the next generation of colliders. Therefore, we expect to either discover the Higgs boson below around 700 GeV or to discover evidence for a strongly interacting electroweak symmetry breaking sector. How this strongly interacting sector might manifest itself is the subject of Chapters 5 and 6.

A further theoretical objection to the Standard Model is that perturbative loop corrections to the Higgs boson squared mass are quadratically divergent and counterterms must be adjusted order by order in perturbation theory to cancel these divergences. If one attempts to embed the Standard Model in a fundamental theory with a high energy scale, then the masses of elementary scalars naturally assume values of order the high energy scale. This is a reflection of the quadratic sensitivity of squared scalar masses to the high-energy scale of the underlying fundamental theory. In contrast, the unitarity arguments above imply that the mass of the Higgs boson of the Standard Model must be of order the electroweak scale. To achieve this requires an unnatural cancelation of terms, each being of order the high energy scale, but whose sum is of order the electroweak scale.

Considering the problems of triviality, perturbative unitarity, and the quadratic divergence in the unrenormalized Higgs boson squared mass leads us to conclude that the Standard Model is at best a good low-energy approximation to a more fundamental theory of particle interactions, valid in a limited domain of "low-energies" of order 1 TeV and below. In light of the many theoretical objections to the simplest version of the Higgs mechanism, theorists have considered several alternatives for electroweak symmetry breaking. In all cases, new physics beyond the Standard Model must arise at or below the TeV scale.

## 4 Model Independent Aspects of New TeV Scale Physics

We begin our discussion of TeV scale physics beyond the Standard Model by considering the features that are independent of the assumptions about the source of the new physics. This is usually accomplished by studying the gauge boson two, three and four-point functions and looking for deviations from the Standard Model values. If we assume that the electroweak gauge group is $SU(2)_L \times U(1)_Y$ and that the scale of new physics is much greater than $m_W$, then the effects of new physics are largest in the corrections to the two-point functions of the gauge bosons. As discussed in Chapter 10, these effects and can be parametrized in terms of three parameters: $S$, $T$, and $U$ (or the related parameters $\epsilon_1$, $\epsilon_2$, and $\epsilon_3$)[10]. The LEP experiments have placed stringent bounds on these parameters which can then be interpreted as bounds on various types of new physics.

It has become conventional in the literature to parametrize the three gauge boson vertices, $\gamma W^+W^-$ and $ZW^+W^-$, in the following manner:

$$\mathcal{L}_{WWV} = -ig\cos\theta_W g_1^Z \left(W_{\mu\nu}^\dagger W^\mu - W_{\mu\nu}W^{\mu\,\dagger}\right)Z^\nu - ieg_1^\gamma \left(W_{\mu\nu}^\dagger W^\mu - W_{\mu\nu}W^{\mu\,\dagger}\right)A^\nu$$

$$-ig\cos\theta_W \kappa_Z W_\mu^\dagger W_\nu Z^{\mu\nu} - ie\kappa_\gamma W_\mu^\dagger W_\nu A^{\mu\nu}$$

$$-g\cos\theta_W g_5^Z \epsilon^{\alpha\beta\mu\nu}\left(W_\nu^- \partial_\alpha W_\beta^+ - W_\beta^+ \partial_\alpha W_\nu^-\right)Z_\mu$$

$$+e\frac{\lambda_\gamma}{m_W^2}W_{\lambda\mu}^\dagger W_\nu^\mu A^{\nu\lambda} + g\cos\theta_W \frac{\lambda_Z}{m_W^2}W_{\lambda\mu}^\dagger W_\nu^\mu Z^{\nu\lambda}\,. \qquad (4.1)$$

This is the most general renormalizable, Lorentz invariant and CP invariant interaction Lagrangian for on-shell gauge bosons [11]. Electromagnetic gauge invariance requires $g_1^\gamma = 1$. In the Standard Model, $g_1^Z = g_1^\gamma = \kappa_Z = 1$ and $g_5^Z = \lambda_\gamma = \lambda_Z = 0$, and within the context of a given model, the coefficients can be predicted. Typically, these coefficients are rather small, of $\mathcal{O}(\alpha/4\pi)$ if they arise, for example, from loop effects. A discussion of the limits that can be placed on $\kappa_Z$, $\kappa_\gamma$, $g_1^Z$, $\lambda_\gamma$, and $\lambda_Z$ at various accelerators is given in Chapter 9. Any measurement of these coefficients that deviates from the Standard Model values would indicate new physics, but could not distinguish the source of the new physics.

A chiral Lagrangian approach is often used to parametrize new physics beyond the Standard Model [12]. To lowest order, this non-linear Lagrangian is simply the Standard Model with no Higgs boson and massive $W$ and $Z$ gauge bosons. The advantage of this approach is that it is a consistent expansion in $s/\Lambda^2$, where $\Lambda$ is the high energy scale that characterizes the scale where new physics occurs. In this picture, the couplings of the two, three and four gauge boson interactions are related. There are five free parameters in the Lagrangian to $\mathcal{O}(s/\Lambda^2)$, assuming an $SU(2)_L \times U(1)_Y$ gauge theory, a custodial SU(2)$_C$ symmetry (which guarantees that the electroweak $\rho$ parameter is equal to 1, as required by experimental data), and CP conservation. One of these parameters (often called $L_{10}$ in the literature) is related to the $S$ parameter measured at LEP, while the other four give non-Standard Model three and four gauge boson couplings. Limits on these couplings are discussed in Chapter 9. [For the three gauge boson couplings, it is straightforward to map the chiral Lagrangian parameters into the parametrization of Eq. (4.1).] It is also possible to couple a generic heavy ($M \sim 1$ TeV) $I = 0$ scalar or $I = 1$ spin-one boson to the chiral Lagrangian and derive valid "low-energy" predictions, even though the system is strongly interacting.

To make further progress, one must consider specific models of the electroweak breaking sector. If the electroweak symmetry breaking sector is weakly coupled then it is straightforward to predict the chiral Lagrangian parameters in terms of the fundamental parameters of the underlying theory. If the electroweak symmetry breaking sector is strongly coupled, this task is far more difficult. We now turn to specific models for the TeV scale physics.

## 5   Low-Energy Supersymmetric Models

Low-energy supersymmetry is a theory of fundamental particle interactions, in which the supersymmetry breaking scale is assumed to be connected with electroweak symmetry breaking (and hence, is assumed to lie between 100 GeV and a few TeV). The simplest version of such a theory is the minimal supersymmetric extension of the Standard Model (MSSM), which consists of taking the Standard Model and adding the corresponding supersymmetric partners [13]. In addition, the MSSM contains two hypercharge $Y = \pm 1$ Higgs doublets, which is the minimal structure for the Higgs sector of an anomaly-free supersymmetric extension of the Standard Model. The supersymmetric structure of the theory also requires (at least) two Higgs doublets to generate mass for both "up"-type and "down"-type quarks (and charged leptons) [14]. All renormalizable supersymmetric interactions consistent with (global) $B - L$ conservation ($B$ = baryon number and $L$ = lepton number) are included. Finally, the most general soft-supersymmetry-breaking terms are added [15]; the mass scale associated with such terms is taken to be of order 1 TeV or below.

As a consequence of $B - L$ invariance, the MSSM possesses a discrete $R$-parity invariance, where $R = (-1)^{3(B-L)+2S}$ for a particle of spin $S$ [16]. Note that this formula implies that all the ordinary Standard Model particles have even $R$-parity, whereas the corresponding supersymmetric partners have odd $R$-parity. The conservation of $R$-parity in scattering and decay processes has a crucial impact on supersymmetric phenomenology. For example, starting from an initial state involving ordinary ($R$-even) particles, it follows that supersymmetric particles must be produced in pairs. In general, these particles are highly unstable and decay quickly into lighter states. However, $R$-parity invariance also implies that the lightest supersymmetric particle (LSP) is absolutely stable, and must eventually be produced at the end of a decay chain initiated by the decay of a heavy unstable supersymmetric particle. In order to be consistent with cosmological constraints, the LSP is almost certainly electrically and color neutral [17]. Consequently, the LSP is weakly-interacting in ordinary matter, *i.e.*, it behaves like a heavy stable neutrino and will escape detectors without being directly observed. Thus, the canonical signature for $R$-parity conserving supersymmetric theories[a] is missing (transverse) energy, due to the escape of the LSP.

The parameters of the MSSM are conveniently described by considering separately the supersymmetry-conserving sector and the supersymmetry-breaking sector. Supersymmetry breaking is accomplished by including the most general set of soft-supersymmetry breaking terms; these terms parametrize our ignorance of the fundamental mechanism of supersymmetry breaking. Among the parameters of the supersymmetry conserving sector are: (i) gauge couplings: $g_s$, $g$, and $g'$, corresponding to the Standard Model gauge group SU(3)×SU(2)×U(1) respectively; (ii) Higgs Yukawa couplings: $\lambda_e$, $\lambda_u$, and $\lambda_d$ (which are $3 \times 3$ matrices in flavor space); and (iii) a supersymmetry-conserving Higgs mass parameter $\mu$. The supersymmetry-breaking sector contains the following set of parameters: (i) gaugino Majorana masses $M_3$, $M_2$ and $M_1$ associated with the $SU(3)_C$, $SU(2)_L$, and $U(1)_Y$ subgroups of the Standard Model; (ii) scalar mass matrices for the squarks and sleptons; (iii) Higgs-squark-squark trilinear interaction terms (the so-called "$A$-parameters") and corresponding terms involving the sleptons; and (iv) three scalar Higgs mass parameters—two diagonal and one off-diagonal mass terms for the two Higgs doublets. These three mass parameters can be re-expressed in terms of the two Higgs vacuum expectation values, $v_1$ and $v_2$, and one physical Higgs mass. Here, $v_1$ ($v_2$) is the vacuum expectation value of the Higgs field which couples exclusively to down-type (up-type) quarks and leptons. Note that $v_1^2 + v_2^2 = (246 \text{ GeV})^2$ is fixed by the $W$ mass, while the ratio

$$\tan \beta = v_2/v_1 \qquad (5.1)$$

is a free parameter of the model.

The supersymmetric constraints imply that the MSSM Higgs sector is automatically CP-conserving (at tree-level). Thus, $\tan \beta$ is a real parameter (conventionally chosen to be positive), and the physical neutral Higgs scalars are CP-eigenstates. Nevertheless, the MSSM does contain a number of possible new sources of CP violation. For example, gaugino mass parameters, the $A$-parameters, and $\mu$ may be complex. Some combination of these complex phases must be less than of order $10^{-2}$–$10^{-3}$ (for a supersymmetry-breaking scale of 100 GeV) to avoid generating electric dipole moments for the neutron, electron, and atoms in conflict with observed data [19]. However, these complex phases have little impact on the direct searches for supersymmetric particles, and are usually ignored in experimental analyses.

---

[a]Some model builders attempt to relax the assumption of $R$-parity conservation. Models of the type must break $B - L$ and are therefore strongly constrained by experiment [18]. In such models, the LSP is unstable and supersymmetric particles can be singly produced and destroyed in association with $B$ or $L$ violation. These features lead to a phenomenology of broken-$R$-parity models that is very different from that of the MSSM.

Before describing the supersymmetric particle sector, let us consider the Higgs sector of the MSSM [20]. There are five physical Higgs particles in this model: a charged Higgs pair ($H^\pm$), two CP-even neutral Higgs bosons (denoted by $h^0$ and $H^0$ where $m_{h^0} \leq m_{H^0}$) and one CP-odd neutral Higgs boson ($A^0$). The properties of the Higgs sector are determined by the Higgs potential which is made up of quadratic terms [whose squared-mass coefficients were mentioned above Eq. (5.1)] and quartic interaction terms. The strengths of the interaction terms are directly related to the gauge couplings by supersymmetry. As a result, $\tan\beta$ [defined in Eq. (5.1)] and one Higgs mass determine: the Higgs spectrum, an angle $\alpha$ (which indicates the amount of mixing of the original $Y = \pm 1$ Higgs doublet states in the physical CP-even scalars), and the Higgs boson couplings. When one-loop radiative corrections are incorporated, additional parameters of the supersymmetric model enter via virtual loops. The impact of these corrections can be significant [21,22]. For example, at tree-level, the MSSM predicts $m_{h^0} \leq m_Z$ [14]. If true, this would imply that experiments to be performed at LEP-2 operating at its maximum energy and luminosity would rule out the MSSM if $h^0$ were not found. However, this Higgs mass bound can be violated when the radiative corrections are incorporated. For example, in [21], the following approximate upper bound was obtained for $m_{h^0}$ (assuming $m_{A^0} > m_Z$) in the limit of $m_Z \ll m_t \ll M_{\tilde{t}}$ [where top-squark ($\tilde{t}_L$–$\tilde{t}_R$) mixing is neglected]

$$m_{h^0}^2 \lesssim m_Z^2 + \frac{3g^2 m_Z^4}{16\pi^2 m_W^2} \left\{ \left[ \frac{2m_t^4 - m_t^2 m_Z^2}{m_Z^4} \right] \ln\left(\frac{M_{\tilde{t}}^2}{m_t^2}\right) + \frac{m_t^2}{3m_Z^2} \right\}. \quad (5.2)$$

More refined computations [23] (which include the effects of top-squark mixing at one-loop, renormalization group improvement, and the leading two-loop contributions) yield $m_{h^0} \lesssim 125$ GeV for $m_t = 175$ GeV and a top-squark mass of $M_{\tilde{t}} = 1$ TeV. Clearly, the radiative corrections to the Higgs masses have a significant impact on the search for the MSSM Higgs bosons at LEP-2 [4].

Consider next the supersymmetric particle sector of the MSSM; further details can be found in Ref. [24]. The *gluino* is the color octet Majorana fermion partner of the gluon with mass $M_{\tilde{g}} = |M_3|$. The supersymmetric partners of the electroweak gauge and Higgs bosons (the *gauginos* and *higgsinos*) can mix. As a result, the physical mass eigenstates are model-dependent linear combinations of these states, called *charginos* and *neutralinos*, which are obtained by diagonalizing the corresponding mass matrices. The chargino mass matrix depends on $M_2$, $\mu$, $\tan\beta$ and $m_W$. The corresponding chargino mass eigenstates are denoted by $\tilde{\chi}_1^+$ and $\tilde{\chi}_2^+$ (some authors use the notation $\widetilde{W}_1^+$ and $\widetilde{W}_2^+$), with masses

$$M^2_{\tilde{\chi}_1^+, \tilde{\chi}_2^+} = \tfrac{1}{2}\left(|\mu|^2 + |M_2|^2 + 2m_W^2\right) \mp \Big\{\left(|\mu|^2 + |M_2|^2 + 2m_W^2\right)^2 \\ - 4|\mu|^2|M_2|^2 - 4m_W^4 \sin^2 2\beta + 8m_W^2 \sin 2\beta\, \mathrm{Re}(\mu M_2)\Big\}^{1/2}, \quad (5.3)$$

where the states are ordered such that $M_{\widetilde{\chi}_1^+} \leq M_{\widetilde{\chi}_2^+}$. If CP-violating effects are ignored (in which case, $M_2$ and $\mu$ are real parameters), then one can choose a convention where $\tan\beta$ and $M_2$ are positive. (Note that the relative sign of $M_2$ and $\mu$ is meaningful. The sign of $\mu$ is convention-dependent; the reader is warned that both sign conventions appear in this book!) The sign convention for $\mu$ implicit in Eq. (5.3) is used by the LEP collaborations [25] in their plots of exclusion contours in the $M_2$ vs. $\mu$ plane derived from the non-observation of $Z \to \widetilde{\chi}_1^+ \widetilde{\chi}_1^-$. The $4 \times 4$ neutralino mass matrix depends on $M_1$, $M_2$, $\mu$, $\tan\beta$, $m_Z$, and the weak mixing angle $\theta_W$. The corresponding neutralino eigenstates are usually denoted by $\widetilde{\chi}_i^0$, $i = 1, \ldots 4$ (some authors use the notation $\widetilde{Z}_i$), according to the convention that $M_{\widetilde{\chi}_1^0} \leq M_{\widetilde{\chi}_2^0} \leq M_{\widetilde{\chi}_3^0} \leq M_{\widetilde{\chi}_4^0}$. Typically, $\widetilde{\chi}_1^0$ is the LSP.

It is common practice in the literature to reduce the supersymmetric parameter freedom by requiring that all three gaugino mass parameters are equal at some grand unification scale. Then, at the electroweak scale, the gaugino mass parameters can be expressed in terms of one of them (say, $M_2$) and the gauge coupling constants:

$$M_3 = (g_s^2/g^2)M_2 \, , \qquad M_1 = (5g'^2/3g^2)M_2 \, . \qquad (5.4)$$

Having made this assumption, the chargino and neutralino masses and mixing angles depend only on three unknown parameters: the gluino mass, $\mu$, and $\tan\beta$. However, the assumption of gaugino mass unification could prove false and must eventually be tested experimentally.

The supersymmetric partners of the quarks and leptons are spin-zero bosons: the *squarks*, charged *sleptons*, and *sneutrinos*. For a given fermion $f$, there are two supersymmetric partners $\widetilde{f}_L$ and $\widetilde{f}_R$ which are scalar partners of the corresponding left and right-handed fermion. (There is no $\widetilde{\nu}_R$.) However, in general, $\widetilde{f}_L$ and $\widetilde{f}_R$ are not mass-eigenstates since there is $\widetilde{f}_L$-$\widetilde{f}_R$ mixing which is proportional in strength to the corresponding element of the scalar squared-mass matrix [26]:

$$M_{LR}^2 = \begin{cases} m_d(A_d - \mu \tan\beta), & \text{for ``down''-type } f \\ m_u(A_u - \mu \cot\beta), & \text{for ``up''-type } f, \end{cases} \qquad (5.5)$$

where $m_d$ ($m_u$) is the mass of the appropriate "down" ("up") type quark or lepton. Here, $A_d$ and $A_u$ are (unknown) soft-supersymmetry-breaking $A$–parameters and $\mu$ and $\tan\beta$ have been defined earlier. The signs of the $A$ parameters are also convention-dependent; see [24]. Due to the appearance of the *fermion* mass in Eq. (5.5), one expects $M_{LR}$ to be small compared to the diagonal squark and slepton masses, with the possible exception of the top-squark, since $m_t$ is large, and the bottom-squark and tau-slepton if $\tan\beta \gg 1$. The (diagonal) $L$ and $R$-type squark and slepton masses are given by [27]

$$\begin{aligned} M^2_{\widetilde{f}_L} &= M^2_F + m^2_f + m^2_Z \cos 2\beta (T_3 - e_f \sin^2\theta_W)\,, \\ M^2_{\widetilde{f}_R} &= M^2_R + m^2_f + e_f m^2_Z \cos 2\beta \sin^2\theta_W\,, \end{aligned} \quad (5.6)$$

where $M_F \equiv M_{\widetilde{Q}}$ [$M_{\widetilde{L}}$] for $L$-type squarks [sleptons], with the appropriate choice of $T_3 = \pm\frac{1}{2}$ for up-type and down-type squarks and sleptons, $M_R = M_{\widetilde{U}}$, $M_{\widetilde{D}}$ and $M_{\widetilde{E}}$ for $\widetilde{u}_R$, $\widetilde{d}_R$ and $\widetilde{e}_R$, respectively, and $m_f$ and $e_f$ are the corresponding quark [lepton] mass and charge (in units of $e$). The soft-supersymmetry-breaking parameters $M_{\widetilde{Q}}$, $M_{\widetilde{U}}$, $M_{\widetilde{D}}$, $M_{\widetilde{L}}$, and $M_{\widetilde{E}}$ are unknown parameters. Note that generational indices have been suppressed; further complications such as intergenerational mixing are possible, although there are some constraints from the nonobservation of flavor-changing neutral currents (FCNC) [28]. One way to guarantee the absence of significant FCNC's mediated by virtual supersymmetric particle exchange is to posit that the diagonal soft-supersymmetry-breaking scalar squared-masses are universal in flavor space at some energy scale (normally taken to be at or near the Planck scale) [29]. Renormalization group evolution is used to determine the low-energy values for the scalar mass parameters listed above. This assumption substantially reduces the MSSM parameter freedom.

Two additional theoretical frameworks are often introduced to reduce further the MSSM parameter freedom [27,30]. The first involves grand unified theories (GUTs) and the desert hypothesis (*i.e.*, no new physics between the TeV-scale and the GUT-scale). Perhaps one of the most compelling hints for low-energy supersymmetry is the unification of SU(3)×SU(2)×U(1) gauge couplings predicted by supersymmetric GUT models [29,31] (with the supersymmetry breaking scale of order 1 TeV or below). The unification, which takes place at an energy scale of order $10^{16}$ GeV, is quite robust (and depends weakly on the details of the GUT-scale theory). In contrast, gauge coupling unification in the simplest nonsupersymmetric GUT fails by many standard deviations [32]. The second framework involves minimal supergravity theory, where the soft supersymmetry-breaking parameters are often taken to have the following simple form. Referring to the parameter list given above Eq. (5.1), the Planck-scale values of the soft-supersymmetry-breaking terms depend on the following minimal set of parameters: (i) a universal gaugino mass $m_{1/2}$; (ii) a universal diagonal scalar mass parameter $m_0$; (iii) a universal $A$-parameter, $A_0$; and (iv) three scalar Higgs mass parameters—two common diagonal squared-masses given by $|\mu_0|^2 + m_0^2$ and an off-diagonal squared-mass given by $B_0\mu_0$ (which defines the Planck-scale supersymmetry-breaking parameter $B_0$), where $\mu_0$ is the Planck-scale value of the $\mu$-parameter. Renormalization group evolution is used to compute the low-energy values of the supersymmetry-breaking parameters and determines the supersymmetric particle spectrum. Moreover, in this approach, electroweak symmetry breaking is induced radiatively if one of the Higgs diagonal squared-masses is forced negative by the evolution. However, the minimal approach above is probably too restrictive. Theoretical considerations suggest that the universality of Planck-scale soft-supersymmetry breaking parameters is not generic [33]. Recent developments are reviewed in Chapter 3.

The verification of low-energy supersymmetry requires the discovery of the supersymmetric particles. Once superpartners are discovered, it is necessary to test their detailed properties to verify the supersymmetric nature of their interactions. Furthermore, one can explicitly test many of the additional theoretical assumptions described above that were introduced to reduce the supersymmetric parameter freedom. The phenomenology of supersymmetric particles is discussed in detail in Chapter 4. The phenomenology of the Higgs sector may also provide crucial evidence for an underlying low-energy supersymmetry. If $m_{A^0} \gg m_{h^0}$, then the properties of $h^0$ will be nearly indistinguishable from the Higgs boson of the minimal Standard Model [34]. Small deviations from the Standard Model Higgs sector could signal the existence of additional Higgs states, as expected in the MSSM. If $m_{A^0} \sim \mathcal{O}(m_Z)$, then the full non-minimal Higgs sector could be revealed soon after the initial discovery of the first Higgs state. The phenomenology of the MSSM Higgs bosons is considered in Chapter 2.

## 6 Models with Dynamical Electroweak Symmetry Breaking

One of the main examples of dynamical electroweak symmetry breaking is the technicolor approach, which postulates the existence of technicolor gauge interactions acting among a set of new, massless technifermions, $\psi_{TF}$ [35]. The technifermions carry the ordinary quantum numbers under SU(3)$_C$×SU(2)$_L$×U(1)$_Y$, plus their own technicolor charges. The Standard Model particles are all singlets under the technicolor gauge group. The technicolor gauge group and technifermion representations are chosen so that the theory possesses a custodial SU(2)$_C$ symmetry, which ensures that $\rho = m_W^2/(m_Z^2 \cos^2\theta_W) = 1$. In the simplest version, there are $N_{TF}$ technifermions which transform in the fundamental **N** representation of the SU(N) technicolor group. The theory thus has a chiral SU(N$_{TF}$)$_L$ × SU(N$_{TF}$)$_R$ global symmetry. The technicolor interaction is chosen to be asymptotically free and to have a particle spectrum such that at an energy scale of $\mathcal{O}(1\ TeV)$ it becomes strong. At this energy scale, one assumes that condensates of technifermions form,

$$\langle \overline{\psi}_{TF} \psi_{TF} \rangle \neq 0 \quad . \tag{6.1}$$

This breaks the chiral symmetry to a vector-like SU(N$_{TF}$) and gives rise to $N_{TF}^2 - 1$ Goldstone bosons called technipions, in much the same way that the light pseudoscalar mesons arise in QCD. Three of the technipions become the longitudinal components of the $W^\pm$ and $Z$ and hence these gauge bosons acquire mass. In order to obtain the observed $W$ and $Z$ masses, the technicolor scale is fixed to be

$$\Lambda_{TC} \sim 4\pi v \sqrt{\frac{2}{N_{TF}}}. \tag{6.2}$$

Since the minimal anomaly free choice of technifermions has eight flavors of technifermions with the Standard Model quantum numbers of a single generation of quarks and leptons, one indeed finds $\Lambda_{TF} \sim 1$ TeV.

After the $W$ and $Z$ get mass (by absorbing three Goldstone technipion states), the remaining technipions are physical particles. The number of such technipions depends on the gauge group chosen for the technicolor interaction and on the representations chosen for the technifermions. The minimal constraints, such as the requirement that the technicolor interaction becomes strong at about a TeV, require a gauge group such as SU(4) or larger and yields a rich spectrum of technipions. Since the technifermions also carry ordinary electroweak and color charges, the technipions occur in multiplets of these groups. Much of the phenomenology of technicolor models involves the search for technipions, which typically have masses at the electroweak scale.

The technicolor spectrum mimics the QCD spectrum and so there are, for example, mesons with the quantum numbers of the $\rho$ and $\omega$. Large $N$ arguments lead to the estimate

$$M_{\rho_{TC}} \sim 2 \text{ TeV} \sqrt{\frac{3}{N}} \tag{6.3}$$

which is yet another indication that the relevant scale for electroweak symmetry breaking is 1 TeV. Extensive effort has been spent predicting experimental signatures for the techni-rho and the techni-omega [36]; see Chapter 6 for further details.

The basic technicolor mechanism leaves the quarks and leptons massless and extended technicolor interactions (ETC) must be introduced to give the fermions mass. There is no version of ETC that naturally gives fermions mass without also generating flavor changing neutral currents (FCNC) in conflict with experimental data; no simple equivalent of the GIM mechanism has been found. The ETC interactions contribute, for example, to the $K_L$–$K_S$ mass difference, which requires $M_{ETC} > 200$ TeV in order to suppress this interaction to an acceptable level.

The simplest technicolor models give fermion masses

$$m_f \sim \frac{\langle \overline{\psi}\psi \rangle}{M_{ETC}^2}, \tag{6.4}$$

where naively $\langle \overline{\psi}\psi \rangle \sim \Lambda_{TC}^3$. Taking $\Lambda_{TC} \sim 1$ TeV and $M_{ETC} \sim 200$ TeV, we see that it is impossible to generate large enough masses for the $s$ and $c$ quarks, let alone the third generation quarks. One solution is to have a hierarchy of scales, with a different ETC scale for each generation. Obtaining the observed top quark mass and generating the observed top–bottom mass splitting is still difficult [37]. The simple picture described above can be avoided in "walking" technicolor models, which change the naive scaling of Eq. (6.4) [38]. If the technicolor model has a large mass anomalous dimension then a larger value of $M_{ETC}$ can be used to obtain the observed quark masses, which in turn will suppress problems with FCNCs. The low ETC scale required to get $m_t \simeq 175$ GeV will lead to a predicted value for the $Z \to b\bar{b}$ branching ratio in conflict with the LEP data in most models of extended technicolor, although this can be avoided in some models where the ETC gauge group does not commute with the electroweak SU(2) group [39].

The large mass of the top quark has led to suggestions that it may be fundamentally different from the five lighter quarks. Another example of dynamical symmetry breaking is a class of models in which the top quark forms a condensate at some high energy scale ($\Lambda_t$)[40], $\langle t\bar{t} \rangle \neq 0$. This approach yields a CP-even neutral scalar bound state whose properties are indistinguishable from those of the Standard Model Higgs boson. In these models, the scalar mass is fixed by the dynamics, and tends to be around 200 GeV. However, this is typically achieved only when $\Lambda_t \gg m_Z$, in which case there is a fine-tuning problem identical to that encountered in the Standard Model. Attempts have been made to incorporate the top quark condensate in a dynamical scheme where fine-tuning is avoided by taking $\Lambda_t \sim \mathcal{O}(1 \text{ TeV})$ and embedding this structure in a theory of extended technicolor [37,41]. Interesting models have been constructed. Nevertheless, the dynamical electroweak symmetry breaking approach suffers from the fact that no simple (and calculable) minimal extension of the Standard Model has been found that satisfies all present experimental constraints.

# 7  Capsule Summary

To summarize, the primary goals of TeV-scale exploration are: to reveal the nature of the Higgs sector and discover the underlying dynamics of electroweak symmetry breaking; to discover whether new particles and interactions beyond the Standard Model exist at the TeV-scale (*e.g.*, confirm or rule out low-energy supersymmetry); and to uncover new insights for physics at even higher energy scales which impact on unification, the origin of flavor, *etc.* Exploration of the TeV-scale requires a new generation of colliders and detectors beyond those currently in operation or construction. This book presents an in-depth study of the phenomenology of electroweak symmetry breaking and quantifies the "physics reach" of present and future colliders. Its focus is on the Standard Model (with one Higgs doublet) and beyond: models of low-energy supersymmetry, dynamical electroweak symmetry breaking, and other approaches that invoke new particles and interactions. We hope that this volume will serve a useful purpose and provide a foundation for developing the future direction of particle physics at the dawn of the 21st century.